\magnification=1200
\baselineskip14pt plus1pt minus1pt
\overfullrule0pt

\font\bff=cmb10
\font\bb=msbm10

\def\R{\hbox{\bb R}}
\def\Z{\hbox{\bb Z}}
\def\bs{\bigskip}
\def\ms{\medskip}
\def\ss{\smallskip}
\def\nt{\noindent}

\def\i#1/#2\par{\item{\hbox to .6truecm{#1\hss}}#2}
\def\ii#1/#2/#3\par{\advance\parindent by .8truecm
       \item{\hbox to .6truecm{#1\hss}
         \hbox to 0.6truecm{#2\hss}}#3
          \advance\parindent by -.8truecm}
\def\iii#1/#2/#3/#4\par{\advance\parindent by 1.6truecm
        \item{\hbox to 0.6truecm{#1\hss}
              \hbox to 0.6truecm{#2\hss}
              \hbox to 0.6truecm{#3\hss}}#4
        \advance\parindent by -1.6truecm}


\centerline{\bff TOPOLOGICALLY INDUCED INSTABILITY IN STRING THEORY} 
\bs
\centerline{Brett McInnes}
\ms
\centerline{Department of Mathematics}
\ss
\centerline{National University of Singapore}
\ss
\centerline{10 Kent Ridge Crescent, Singapore 119260}
\ss
\centerline{Republic of Singapore}
\bs\bs\bs

\nt{\bf ABSTRACT:} \ \ Using Witten's generalised $AdS/CFT$ correspondence, we show that there are certain differentiable manifolds $Z^{10}$ such that string theory is unstable [to the emission of ``large branes''] on $Z^{10}$ {\it no matter what the metric may be}.  Therefore, the instability is due to the [differential] topology of the background, not to its geometry.  We propose a precise formulation of this restriction on the topology of string backgrounds, and give evidence for its validity.  The discussion is designed to introduce particular geometric/topological techniques which are likely to be useful in further applications of the $AdS/CFT$ correspondence.

\bs\bs

\nt{\bf 1. INTRODUCTION}
\ss
One of the drawbacks of General Relativity is that it fixes the topology of spacetime, if at all, only partially and indirectly [1].  For example, should it prove to be the case that our four-dimensional Universe can be represented by a FRW geometry with {\it compact} spatial sections of constant negative curvature [2], then we cannot expect GR to explain why the fundamental group of the sections has one structure rather than another: for that, we must look to string theory.

\ss

Unfortunately, the precise way in which string theory fixes topology is far indeed from being understood.  In order to study this question, we need concrete examples of string backgrounds such that the theory misbehaves in some way, {\it no matter what the metric tensor may be}, so that the misbehaviour can be attributed to the {\it topology} of the space in question rather than to its {\it geometry}.  A precise formulation of the topological property responsible for the misbehaviour might throw some light on the general ways in which string theory eliminates some topologies in favour of others.

\ss

Seiberg and Witten [3] have studied a fundamental kind of potential ``misbehaviour'' in string theory in the context of Type IIB on $AdS_5 \times S^5$.  They observe that this system is 
potentially unstable against the emission of ``large'' 3-branes. The relevant action has the
following form:

$$S = \cases{ {Tr^D_0 \over 2^D} \int \sqrt{g}\; \left( (1-q)\phi^{{2D\over D-2}}+{8 \over (D-2)^2}[(\partial\phi)^2 + {D-2\over 4(D-1)}\phi^2R] + {\cal O}(\phi^{{2(D-4)\over D-2}})\right) &for $D > 2$; \cr
{Tr^2_0 \over 4} \int \sqrt{g}\; ((1-q)e^{2\phi} + 2[(\partial \phi)^2 + \phi R] - R + {\cal O}(e^{-2\phi})) &for $D = 2$. \cr}$$

Here we have written the general action for a [D-1]-brane carrying a charge
$q$ under a background antisymmetric field; the field ${\phi}$ tends to
infinity as the conformal boundary is approached, and $R$ is the scalar
curvature of that boundary. There is clearly a danger that the action
will be unbounded below if either $q$ exceeds 1 or $R$ is negative; in
such cases the system is unstable. The first case, however, never arises
in a supersymmetric background, because it would violate a BPS bound [3].
The second possibility does not occur in the $AdS$ case, because the 
conformal structure at infinity is represented by a metric of {\it positive}
scalar curvature. In short, this "Seiberg-Witten brane instability" 
problem does not arise in ordinary $AdS/CFT$ because supersymmetry takes
care of values of $q$ larger than 1, and the positivity of the scalar
curvature on the boundary takes care of the case when $q = 1$, that is,
the case of BPS branes.

\ss

Conversely, it is also observed in [3] that if $AdS_5$ is replaced by a more general non-compact negatively curved manifold, then the scalar curvature of the conformal boundary may be negative, so that the system becomes unstable to the emission of large BPS branes.  [The crucial point here is that, even if the more general metric is {\it asymptotically} the same as the Euclidean $AdS$ metric, the geometry it induces on the conformal boundary can be very different to that induced by the usual AdS metric.  See the examples given in Section 2, below.]

\ss

According to the $AdS/CFT$ correspondence, as generalised in [4], this phenomenon should have a conformal field theory counterpart, and so it does [3], [5], [6] : the particular kind of CFT which arises in $AdS/CFT$ is {\it stable if the scalar curvature of the space on which it is defined is positive, and sometimes when the scalar curvature is zero, but it is unstable when the scalar curvature is negative}.  [We follow these references and use Euclidean signature throughout.]  The generalised $AdS/CFT$ correspondence of [4] allows us to detect Seiberg-Witten string instability by studying the geometry at infinity.

\ss

Now suppose we consider the theory of $(n-1)$-branes on $W^{n+1} \times Y^{9-n}$, where $W^{n+1}$ is an $(n+1)$-dimensional non-compact asymptotically hyperbolic manifold [generalising $AdS_{n+1}$] and where $Y^{9-n}$ is compact, generalising $S^{9-n}$. [A non-compact Riemannian manifold is said to be asymptotically hyperbolic if it resembles a space of constant negative curvature [Euclidean $AdS$] near infinity.  A precise definition will be given later.]  Suppose also that $W^{n+1}$ is such that string theory on $W^{n+1} \times Y^{9-n}$ is stable against the emission of large branes.  Thinking of  $W^{n+1} \times Y^{9-n}$ as a ``vacuum'', consider now the consequences of {\it deforming the geometry} of this space [that is, changing the metric but not the topology].  Physically, this might correspond to the back-reaction from exciting some matter fields in the bulk.  Is it always possible to {\it destabilise} the theory in this way?  Conversely, and more importantly here, is it always possible to {\it stabilise} an unstable string configuration by deforming the geometry?
This question is crucial in any discussion of brane instability, because
eventually one might suspect that the back-reaction from such a process
might change the geometry, so that the instability could be self-limiting.
{\it This would happen if the back-reaction could eventually cause the scalar curvature of conformal infinity to change sign.}

\ss

  The main objective of this work is to show that these questions, despite their difficulty, can often be answered definitively.  This is possible because the generalised $AdS/CFT$ correspondence reduces them to questions about scalar curvature on compact manifolds, a field of geometry in which very powerful mathematical techniques are available.  [Our other objective is to explain these techniques in a way that should be useful to string theorists.]

\ss

Begin with the case in which the initial ``vacuum'' geometry on $W^{n+1}$ is such that the theory is stable in the Seiberg-Witten sense.  Our objective is to determine, by distorting the metric, whether this stability is to be attributed to the {\it geometry} of this vacuum or to the {\it topology} of the underlying manifold.  Physically, it is reasonable to confine the distortions to those which maintain the asymptotic hyperbolicity [so that infinity remains at infinity, and we can continue to use the generalised $AdS/CFT$ correspondence.]  Otherwise they shall be arbitrary.  According to $AdS/CFT$, the scalar curvature of the conformal boundary is positive or zero.  Our question therefore becomes : can a positive or zero scalar curvature metric $-$ say, the standard round metric on $S^4-$ be deformed to such an extent that the scalar curvature becomes negative everywhere?  Bizarre as it may seem, {\it this is indeed always possible} on any compact manifold of positive or zero scalar curvature of dimension at least 3 [see [7], page 123, or section 2 below.]  Therefore, a string theory which is stable against the emission of large branes can always be made unstable by deforming the geometry of the background [provided the dimension of the conformal boundary is at least 3, the only case we shall consider here.]  Thus, {\it stability} is never ``topological'' in the sense we mean here.

\ss

Apart from the fact that we were able to prove it, this assertion is not very surprising, since the brane action does of course depend on the metric.  The answer to the second question above is, however, much more startling.  In each dimension above 1, there exist compact manifolds on which it is {\it impossible to define a Riemannian metric of positive or zero scalar curvature}.  String theory on the $AdS/CFT$ dual ``vacuum'' space will therefore be unstable to the emission of  large branes, {\it and will remain unstable no matter how the metric is distorted}.  In other words, the instability is due not to the kinds of matter fields excited in the bulk or to the choice of metric, but rather to the structure of the underlying manifold itself : we shall speak of {\it topologically induced instability}.  Presumably, string theory forbids such topologies.

\ss

The existence of topologically induced instability provides an interesting and physically well-motivated example of string theory imposing a condition directly on the topology of space-time.  For example, let $N^4$ be a compact four-dimensional hyperbolic manifold [8] [that is, we take the usual simply connected space of constant negative curvature and perform topological identifications so that the quotient is compact].  It is possible to prove that $N^4$ is the boundary of some manifold-with-boundary ${\overline W}^5$ [note that this is not true of all  four-dimensional compact manifolds] and that there is {\it no} metric of positive or zero scalar curvature on $N^4$ [see [9], page 306].  Hence, string theory on $W^5 \times Y^5$, where $W^5$ is the interior of ${\overline W}^5$ and $Y^5$ is any compact five-manifold, is intrinsically unstable to the emission of large branes $-$ ``intrinsically'' in the sense that the conclusion holds true no matter which metric is used.  String theory {\it forbids a background with this topology}.  [See [10] and [11] for other string backgrounds involving compact hyperbolic spaces in a different way.  Note that one can prove [see below] that there exists on $W^5$ an asymptotically hyperbolic metric such that the boundary is at infinity and such that the canonical conformal structure is induced on the boundary.]

\ss

If topologically induced instability is to be useful as a first step towards understanding the way string theory controls space-time topology, we must formulate it precisely.  That is, we must give an unambiguous description of the class of manifolds on which string theory misbehaves in the above way.  Constructing such a description is the task of this work.  The description is surprisingly subtle, so much so indeed that it requires us to be more precise in our usage of the term ``topological''.  To see that the problem is highly non-trivial, recall the importance of positive scalar curvature on the boundary, and take note of the following extraordinary fact : in dimensions 9 and 10, {\it every} compact spin manifold is homeomorphic [but not necessarily {\it diffeomorphic}] to a manifold which admits no metric of positive scalar curvature [Reference [9], page 298].  On the other hand, every simply connected compact {\it non-spin} manifold of dimension at least 5 admits a metric of positive scalar curvature [ [9], page 299].  Thus, the question of stability is connected [particularly in the physically interesting dimensions 9 and 10] with such arcane matters as the choice of a differentiable structure on a manifold or the existence of a spin structure.  [For example, the sphere $S^9$ does of course admit a stable $CFT$ of the relevant kind if we choose the standard differentiable structure, but this is no longer true if another differentiable structure is chosen.]  To be precise, then, we should speak of ``[differential]-topologically induced instability.''

\ss

Kazdan and Warner [12] have classified compact differentiable manifolds [that is, manifolds with a specific, {\it fixed} differentiable structure] of dimension at least 3 into three broad classes, to be defined below.  We propose that {\it the background}[{\it differential}] {\it topologies forbidden by instability to the emission of large branes are those of the local form} $W^{n+1} \times Y^{9-n}$, {\it where} $W^{n+1}$ {\it is dual to a compact differentiable manifold which belongs to Kazdan-Warner class} $N$.  This is actually a rather strong restriction, because class $N$ is by no means small.  We can prove this proposal [given the truth of $AdS/CFT$ duality and of a very plausible geometric conjecture] for spin manifolds $W^{n+1}$ with $n < 8$; beyond that, we can still give a description of the nature of the potential counter-examples, in the unlikely event that any exist.

\ss

We do not, of course, claim that this is the only or the most important restriction imposed by string theory on the [differential] topology of space-time : instability due to the emission of large branes is undoubtedly just one of many topology selection mechanisms in string theory.  However, the importance of {\it scalar curvature} in constructing conformal Laplacian operators does suggest that the techniques we use will be of value in the search for further examples.

\bs

\nt{\bf 2. \ PRECISE STATEMENT OF THE RESTRICTION}.
\ss

Let $N^n$ be a compact differentiable manifold [again, note that we take this to mean that a differentiable structure is fixed] of dimension $n \ge 3$, and let $f$ be a smooth function on $N^n$.  We say that $f$ is a scalar curvature if there exists a smooth Riemannian metric on $N^n$ having $f$ as its scalar curvature.  On some differentiable manifolds, {\it every smooth} $f$ is a scalar curvature.  Such manifolds are said to be in Kazdan-Warner [12] class $P$ [we follow the terminology of Futaki [13].]  All other manifolds are either in class $Z$ [meaning that $f$ is a scalar curvature if and only if it is either strictly negative at some point, or identically zero] or in class $N$ [meaning that $f$ is a scalar curvature if and only if it is strictly negative at some point].  This classification is differential-topological in nature : for example, $S^9$ can be in $P$ with one choice of differentiable structure, in $N$ with another.  [When we do not specify the differentiable structure, the reader may take it that we choose a canonical structure.]  Compact hyperbolic manifolds are in $N$, as indeed are all compact manifolds of strictly negative [not necessarily constant] sectional curvature [[see [9], page 306]. A very remarkable example of a manifold in $N$ is given by the connected sum of $\R P^7 \times S^2$ with an exotic nine-sphere [see [9], page 301].  This manifold has $\pi_1 = \Z_2$, the universal cover being $S^7 \times S^2$ endowed with the standard differentiable structure, so that $S^7 \times S^2$ is in Kazdan-Warner class $P$.  However, {\it none} of the metrics of positive or zero scalar curvature on $S^7 \times S^2$ is invariant with respect to $\Z_2$.  Therefore, a stable $CFT$ on $S^7 \times S^2$ does not descend to a stable $CFT$ on $[\R P^7 \times S^2] \# [{\hbox{~exotic~}} S^9$], since none of the metrics on $S^7 \times S^2$ which ensure stability survives the projection.  This example shows that factoring by a finite group cannot be ignored in this subject; certainly we must {\it not} assume that our manifolds are simply connected.  Compact flat manifolds are in $Z$ [again, [9], page 306], as is $K3$, and so are all compact 8-dimensional manifolds of linear holonomy $SU(4)$.  Compact 6-dimensional simply connected manifolds of holonomy $SU(3)$, however, are all in $P$, {\it not} $Z$.  [The ``simply connected'' condition here can almost certainly be dropped.]  There is an extensive and highly sophisticated literature [for an introduction, see [14]] relevant to the problem of determining the Kazdan-Warner class of a given differentiable manifold.

\ss

The fact that {\it all} compact manifolds of dimension at least 3 fall into one of the three classes immediately gives us the result, mentioned earlier, that all such manifolds admit a metric of constant negative scalar curvature.  Thus, the $CFT$ in the $AdS/CFT$ correspondence can always be destabilised by modifying the metric.  However, a manifold in class $N$ cannot admit a  metric of positive or zero scalar curvature, so the reverse is {\it not} always possible.  [At this point we should perhaps clarify : as is appropriate for a conformal field theory, the $CFT$ is defined on a manifold ``at infinity'' which only has a Riemannian metric defined modulo arbitrary conformal factors.  A conformal equivalence class can always [15] be represented by a  metric of {\it constant} scalar curvature, and it is to this constant that we refer when we speak of the ``sign'' of the scalar curvature.  A metric on a manifold in class $N$ can be deformed so that the scalar curvature becomes positive over most [not all] of the manifold, but the corresponding metric of constant scalar curvature is such that this constant is nevertheless strictly {\it negative}.]

\ss

Our suggestion is that string theory, through the Seiberg-Witten brane instability mechanism, forbids all those background topologies which are ``dual'' [in the $AdS/CFT$ sense] to a compact manifold $N^n$ in Kazdan-Warner class $N$, {\it and no others}.  In order to investigate this idea, we need a precise formulation of the geometry and topology of the generalised $AdS/CFT$ correspondence itself.

\ss

The $AdS/CFT$ duality is usually formulated in terms of a space of the form ${\overline W}^{n+1} \times Y^{9-n}$, where ${\overline W}^{n+1}$ is a compact manifold-with-boundary and $Y^{9-n}$ is a compact manifold; the ``$CFT$'' space is the boundary of ${\overline W}^{n+1}$, and the ``bulk'' is $W^{n+1} \times Y^{9-n}$ where $W^{n+1}$ is the interior of ${\overline W}^{n+1}$.  It was pointed out in [4] that this formulation is overly restrictive, in at least two ways.  First, one might be interested in studying gauge theory on a compact manifold which is not the boundary of {\it any} manifold-with-boundary; second, one might be interested in studying string theory on a bulk which has the structure $W^{n+1} \times Y^{9-n}$ only ``near infinity'', not globally.  It was also observed in [4] that, in dealing with these objections, we may find it necessary to allow for the geometric consequences of ``branes or stringy impurities of some kind in the bulk''.  We now explain briefly how to deal with these observations.  We begin by assuming that the bulk is indeed a global product, only relaxing this assumption after we have dealt with the ``boundary'' problem.

\ss

First, note that the boundary of a compact manifold-with-boundary ${\overline W}^{n+1}$ can always be regarded as a {\it hypersurface} in a compact manifold ${\hat W}^{n+1}$ : just identify two copies of ${\overline W}^{n+1}$ along the boundary.  The bulk $W^{n+1}$ is then a connected component of the complement of the hypersurface.  This suggests [16] a natural generalisation : simply take $N^n$, the compact ``infinity'' manifold, to be a hypersurface in a more general compact manifold ${\hat W}^{n+1}$.  This  removes the superfluous assumption that $N^n$ must be a boundary.  To see how to formulate the idea that $N^n$ is ``at infinity'', let $W^{n+1}$ be the complement, and let $g^W$ be a Riemannian metric on $W^{n+1}$ with the following properties :

\ss

\i (i)/ $g^W$ is smooth near infinity but not necessarily deep in the bulk.  This is to allow for the presence of physically meaningful singularities such as Randall-Sundrum type branes in the bulk; these are associated with non-smooth metrics [17], and represent the ``stringy impurities'' [4] mentioned earlier.

\ss

\i (ii)/ $g^W$ induces a conformal structure on $N^n$.  This means that there exists a piecewise smooth function $F$ on ${\hat W}^{n+1}$ such that $N^n$ is the zero locus of $F$, $F$ is smooth on $N^n$, $dF \not= 0$ on $N^n$, and $F^2g^W$ extends continuously to a metric $g^{\hat W}_F$ on an entire neighbourhood of $N^n$ in ${\hat W}^{n+1}$.  Clearly, if $F$ exists, then $N^n$ is ``infinitely far'' from all points in $W^{n+1}$; but if $F$ exists it is not unique, so $g^W$ does not induce a unique metric on $N^n$.  Instead it induces, through $g^{\hat W}_F$, a conformal structure there.

\ss

\i (iii)/ $g^W$ is {\it asymptotically hyperbolic}.  Let $\gamma$ be a geodesic in $W^{n+1}$ which approaches $N^n$ [that is, the distance from $\gamma(t)$ to $N^n$, as measured by any $g^{\hat W}_F$, tends to zero as $t \to \infty$].  It can be shown [18] that the sectional curvatures of $g^W$ all approach a common [negative] value along $\gamma$ as $t \to \infty$.  [Of course, [18] is concerned with the ``boundary'' case, but a precisely analogous argument applies here.]  If this negative number does not depend on the point in $N^n$ approached by $\gamma$, then $W^{n+1}$ resembles ``Euclidean $AdS_{n+1}$'' [that is, a space of constant negative sectional curvature] near infinity, and we shall say that it is asymptotically hyperbolic.  This is the appropriate [19] weakening of the $AdS$ condition on the bulk geometry.  [Since the sectional curvatures {\it all} approach the same value along $\gamma$, an ``asymptotically Einstein'' condition would in fact be equivalent to asymptotic hyperbolicity.]  Notice that the condition of asymptotic hyperbolicity does not involve the geometry of $N^n$; the latter can have any sign of curvature even though the bulk curvatures all tend to a negative number.

\ss

We now have the generalised $AdS/CFT$ correspondence [4] formulated as a duality between gauge theory on a compact conformal manifold $N^n$ and string theory on a background of the form $W^{n+1} \times Y^{9-n}$, where $W^{n+1}$ is [a connected component of] the complement of $N^n$ [regarded as a hypersurface in a compact manifold ${\hat W}^{n+1}$], and where $W^{n+1}$ is equipped with a piecewise smooth, asymptotically hyperbolic metric which puts $N^n$ ``at infinity'' and induces the given conformal structure there.  [We have not imposed any requirement on $Y^{9-n}$ other than compactness, but, since $Y^{9-n}$ generalises the sphere $S^{9-n}$, it is probably appropriate to require it to belong to Kazdan-Warner class $P$.  When $n > 6$, we take this to mean that $Y^{9-n} = S^{9-n}$.]

\ss

Finally, let us take up the suggestion that the bulk need only be a product ``near infinity''.  Consider a compact 10-dimensional [or 11-dimensional for $M$ theory] manifold ${\hat Z}^{10}$ which contains a compact hypersurface of the form $N^n \times Y^{9-n}$, $2 < n < 10$, where $Y^{9-n}$ is in Kazdan-Warner class $P$ [which implies, incidentally, that $N^n \times Y^{9-n}$ is likewise in $P$ [except of course if $n=8$ or 9], whatever the class of $N^n$ may be].  It is always possible [using, for example, a tubular neighbourhood [20]] to find an open neighbourhood of $N^n$ in ${\hat Z}^{10}$ of the form ${\hat W}^{n+1} \times Y^{9-n}$, where ${\hat W}^{n+1}$ contains $N^n$.  Let $W^{n+1}$ be the complement of $N^n$ in ${\hat W}^{n+1}$.  Now suppose that there exists a piecewise smooth Riemannian metric $g^Z$ on $Z^{10}$, the complement of $N^n \times Y^{9-n}$ in ${\hat Z}^{10}$, such that it is possible to find a neighbourhood ${\hat W}^{n+1} \times Y^{9-n}$ so that, on $W^{n+1} \times Y^{9-n}$, $g^Z = g^W + g^Y$, where $g^Y$ is a metric of positive scalar curvature [$n\not= 8,9$] on $Y^{9-n}$ and $g^W$ is an asymptotically hyperbolic metric on $W^{n+1}$ which puts $N^n$ at infinity.  [That is, one can find a function $F$ on ${\hat W}^{n+1}$, satisfying the usual conditions, such that $g^Y + F^2g^W$ extends continuously to a metric on all of ${\hat Z}^{10}$.]  This is a very substantial generalisation of the case considered earlier, in which ${\hat Z}^{10} = {\hat W}^{n+1} \times Y^{9-n}$ globally.  Again, the duality [4] is between a gauge theory on $N^n$ and a string theory on $Z^{10}$, the complement of $N^n \times Y^{9-n}$ in ${\hat Z}^{10}$.

\ss

Given $N^n$, we must not expect to find a {\it unique} ${\hat Z}^{10}$ [though {\it some} such ${\hat Z}^{10}$ always exists, as we shall see].  If $N^n$ is physically acceptable, we deal with this by asserting [4] that the $CFT$ partition function is expressed as a  [suitably weighted] {\it sum} of string theory partition functions contributed by the various candidates for $Z^{10}$.  Conversely, we assert that if $N^n$ is such that the $CFT$ necessarily misbehaves, then all of the corresponding $Z^{10}$ candidates must be ruled out.

\ss

We can now state our restriction on the [differential] topology of string backgrounds.  Let $Z^{10}$ be a connected component of the complement, in a compact 10-dimensional manifold ${\hat Z}^{10}$, of a hypersurface of the form $N^n \times Y^{9-n}$, $Y^{9-n}$ being a compact manifold in Kazdan-Warner class $P$.  {\it Then string theory on} $Z^{10}$ {\it is subject to topologically induced instability to the emission of large branes if  and only if $N^n$ belongs to Kazdan-Warner class} $N$.

\ss

A concrete example may be helpful.  Let $N^4$ be a compact 4-dimensional manifold of constant negative curvature $-1/L^2$, with metric $g^N_{ij} \; dx^i \otimes dx^j$, and let $Y^5$ be the sphere $S^5$ with its  usual ``round'' metric $g^Y_{ab} \; dy^a \otimes dy^b$ of curvature $+1/L^2$.

\ss

\nt Define
$${\hat Z}^{10} = S^1 \times N^4 \times Y^5 = {\hat W}^5 \times Y^5, \eqno{(2.1)}$$

\nt where $S^1$ is a circle parametrised by $\theta$ running from $-\pi$ to $+\pi$, and set
$$\eqalignno{
g^Z &= {\rm cosec}^2({\theta \over 2})\left[ {L^2 \over 4} d\theta \otimes d\theta + g^N_{ij} dx^i \otimes dx^j\right] + g^Y_{ab} dy^a \otimes dy^b &{(2.2)} \cr
&= g^W + g^Y. \cr}$$

\nt Then a copy of $N^4 \times Y^5$ resides at $\theta = 0$, and $g^Z$ is well-defined on the complement, $Z^{10}$.  The metric $g^W$ induces a conformal structure on $N^4$ : we simply choose $F = \sin ({\theta \over 2})$.  To see that it is asymptotically hyperbolic, let $\gamma$ be a geodesic which approaches the point $x$ in $N^4$.  Then [again adapting the results of [18]] the sectional curvatures along $\gamma$ all tend to the value
$$K_\infty(x) = -|dF(x)|^2_F, \eqno{(2.3)}$$

\nt where the norm is taken with respect to $F^2 g^W$.  [By ``l'Hopital's rule'', the two occurrences of $F$ in this formula ``cancel'', so that $K_\infty(x)$ is independent of the choice of $F$.]  Thus, $g^W$ is asymptotically hyperbolic if it is possible to choose $F$ so that $^-|dF(x)|^2_F$ is independent of $x$.  Choosing $F = \sin(\theta /2)$, we have $dF (\theta = 0) = {1\over 2} d\theta$, and since 

\nt $\sin^2({\theta \over 2}) g^W = {L^2 \over 4} d\theta \otimes d\theta + g^N_{ij} \; dx^i \otimes dx^j$, we clearly have
$$K_\infty(x) = -|{1\over 2} d\theta|^2_F = -{1\over L^2}, \eqno{(2.4)}$$

\nt so $g^W$ is indeed asymptotically hyperbolic.  [Note that, in proving this, we did not use any property of $g^N$; we only used the fact that $\sin^2({\theta \over 2})$ depends only on $\theta$, together with the observation that the coefficient of $d\theta \otimes d\theta$ in $\sin^2({\theta \over 2})g^W$  is constant.  The fact that $N^4$ has constant curvature $-{1\over L^2}$ is not relevant here.]  Actually, $g^Z$ is a smooth, complete metric.  It can be shown [16] that $g^W$ is an Einstein metric with Ricci curvature
$${\rm Ric}(g^W) = -{4 \over L^2} g^W. \eqno{(2.5)}$$

\nt [To prove this, one only needs the fact that $g^N$ satisfies ${\rm Ric}(g^N) = -{3\over L^2} g^N$, not the fact that the curvature of $N^4$ is constant.]

\ss

\nt By construction, $g^Z$ induces on $N^4$ the conformal structure represented by $g^N$, which of course has negative scalar curvature.  The $CFT$ there is therefore unstable.  The geometry we have constructed on $Z^{10}$ allows us to formulate the generalised $AdS/CFT$ correspondence [4] in this case, and the latter predicts that string theory on this background will be unstable in the Seiberg-Witten sense [3].  But now suppose that we deform $g^Z$, while keeping $N^4$ at infinity.  [We keep $N^4$ at infinity, because otherwise it would not make sense to speak of ``doing string theory on $Z^{10}$''; instead we would be studying string theory on the {\it compact} manifold ${\hat Z}^{10}$.]  The new metric will induce a different conformal structure on $N^4$; yet, no matter how thoroughly we change $g^Z$, the conformal structure on $N^4$ {\it will always be represented by a metric of negative scalar curvature}.  For $N^4$ is in Kazdan-Warner class $N$ : it simply cannot admit a metric of positive or zero scalar curvature.  The generalised $AdS/CFT$ correspondence predicts, therefore, that the string theory on $Z^{10}$ will remain unstable no matter what we do to the metric.  String theory forbids this topology.  [Note that this space is not the same as the one we excluded earlier by regarding $N^4$ as a {\it boundary}.]

\ss

More generally, instead of requiring $N^4$ to be a space of constant negative curvature, we can take it to be {\it any} member of Kazdan-Warner class $N$.  Defining ${\hat Z}^{10}$ as in (2.1), we can again construct the metric (2.2) on $Z^{10}$; then $g^W$ is still asymptotically hyperbolic, and the generalised $AdS/CFT$ correspondence can be set up; again, string theory on $Z^{10}$ suffers topologically induced instability.  [If $N^4$ admits an Einstein metric $g^N$ $-$ this will usually, but not invariably, be the case $-$ then $g^N$ can be normalised to satisfy ${\rm Ric}(g^N) = -{3\over L^2} g^N$, and then $g^W$ will satisfy (2.5).  Note, however, that we are not entitled to demand that the infinity submanifold {\it must} admit an Einstein metric, even when it is in Kazdan-Warner class $P$ : for example, the boundary manifold $S^1 \times S^3$ certainly admits no Einstein metric $-$ see [7], page 161.]

\ss

More generally again, let ${\hat Z}^{10}$ by {\it any} compact 10-dimensional manifold which contains a submanifold $N^n \times Y^{9-n}$, with $N^n$ in Kazdan-Warner class $N$ and $Y^{9-n}$ in class $P$.  Let $(-\epsilon,\epsilon) \times N^n \times Y^{9-n}$ be a tubular neighbourhood of $N^n \times Y^{9-n}$ in ${\hat Z}^{10}$, and let $\theta$ be confined to $(-\epsilon, \epsilon)$, taking $\epsilon < \pi$.  Define a metric on the complement of $\theta = 0$ by (2.2), where $g^N$ and $g^Y$ are suitable metrics on $N^n$ and $Y^{9-n}$.  Extend to any piecewise smooth metric $g^Z$ on $Z^{10}$, the complement of $N^n \times Y^{9-n}$ in ${\hat Z}^{10}$.  Again, the generalised $AdS/CFT$ correspondence can be set up, and string theory on $Z^{10}$ will experience topologically induced instability.

\ss

Thus we see that string theory on $Z^{10}$ undergoes topologically induced instability whenever $Z^{10}$ can be expressed as the complement of a hypersurface of the form $N^n \times Y^{9-n}$, where $N^n$ is in Kazdan-Warner class $N$.  It remains to be seen, however, whether $N^n$ {\it must} be in class $N$ for this to happen.  Suppose that $N^n$ is in class $P$.  Let $g^N$ be a metric on $N^n$ [an Einstein metric if possible] with {\it positive} scalar curvature.  Then on the complement of $N^n \times Y^{9-n}$ in a tubular neighbourhood, define
$$g^Z = {\rm cosech}^2({\theta \over 2})\left[ {L^2\over 4} d\theta \otimes d\theta + g^N_{ij} dx^i \otimes dx^j\right] + g^Y_{ab} \; dy^a \otimes dy^b, \eqno{(2.6)}$$

\nt and extend to a piecewise smooth metric on $Z^{10}$.  Then $g^W$, the first part of (2.6), is asymptotically hyperbolic [and $g^W$ satisfies ${\rm Ric}(g^W) = -{n\over L^2} g^W$ if $g^N$ is Einstein.]  Thus $AdS/CFT$ can be set up in this case, and of course the string theory on $Z^{10}$ with this metric is stable since the conformal structure at infinity is represented by a metric of positive scalar curvature.  [Notice that if ${\hat Z}^{10}$ happens to be $S^1 \times N^n \times Y^{9-n}$ [globally] then the extension of (2.6) to $Z^{10}$ is indeed {\it piecewise} smooth, since ${\rm cosech}^2 ({\theta \over 2})$, unlike ${\rm cosec}^2({\theta \over 2})$, does not have zero derivative at $\theta = \pm \pi$.  This is a general fact when the conformal structure at infinity is represented by a metric of positive or zero scalar curvature; for the possible physical significance of this, see [17].]  Thus we see that, as predicted, topologically induced instability cannot occur when $N^n$ is in Kazdan-Warner class $P$.

\ss

There remains the possibility that topologically induced instability could arise when $N^n$ is in Kazdan-Warner class $Z$.  Again, the generalised $AdS/CFT$ correspondence can be set up : choose a scalar-flat metric $g^N$ on $N^n$, and replace (2.6) by
$$g^Z = \theta^{-2} [L^2 d\theta \otimes d\theta + g^N_{ij} \; dx^i \otimes dx^j] + g^Y_{ab} \; dy^a \otimes dy^b, \eqno{(2.7)}$$

\nt extending to $Z^{10}$ as usual.  Here $g^W$, the first part of the formula, is clearly asymptotically hyperbolic, and in fact it is {\it necessarily} an Einstein metric, satisfying ${\rm Ric}(g^W) = - {n\over L^2} g^W$, since $g^N$  must be Ricci-flat [see below].

\ss

But while $AdS/CFT$ can thus be set up when $N^n$ is in class $Z$, it is not clear what it tells us.  Although scalar-flat metrics certainly exist on such manifolds, the vanishing of the scalar curvature means that stability of the conformal field theory is determined by very complex higher-order terms [involving combinations of higher curvature invariants] in the expansion of the relevant action [5].  A stability criterion could be formulated in this way, but it would be pointless to do so : the metrics in question are Ricci-flat, and so [except for flat manifolds] they are not known explicitly, and so the stability condition cannot be verified unless we have further information on the manifold.  Hence we cannot determine in this way whether there exist $N^n$ in Kazdan-Warner class $Z$ such that string theory on $Z^{10}$ is afflicted with topologically induced instability.

\ss

Happily, class $Z$ is, in a sense, much ``smaller'' than classes $P$ or $N$.  In fact, given the truth of an exceedingly plausible geometric conjecture, we can actually list the members of class $Z$ rather explicitly in low dimensions.  This is the objective of the next section.  We shall then return to the stability question in Section 4.                                                                                                                                                                                                                                                                                                                                                                                                                                                                                                                                                                                                                                                                                                                                                                                                                                                                                                                                                                                                                                                                                                                                                                                                                                                                                                                                                                                                                                                                                                                                                                                                                                                                                                                                        

\bs

\nt{\bf 3. \ LOW-DIMENSIONAL MANIFOLDS IN KAZDAN-WARNER CLASS Z}.
\ss
\nt Recall that compact differentiable manifolds in Kazdan-Warner class $Z$, like those in class $N$, admit metrics of constant negative scalar curvature, and fail to admit any metric of constant positive scalar curvature; unlike them, however, manifolds in $Z$ do always admit a metric of zero scalar curvature.  These manifolds therefore play a critical role in the study of the stability of Maldacena-type conformal field theories.  Typical examples of such manifolds are the tori $T^r$ and $K3$.  [While there is a very extensive literature devoted to class $N$, the same cannot be said of class $Z$.  See [13], [21], [22], but note that these works do not study the low-dimensional cases with which we are primarily concerned here.]

\ss

Class $Z$ has the following elementary but interesting property.

\ss

\nt {\bf THEOREM (3.1) [``DIVISION''] :} \ \  Let $M_1$ be a compact manifold which admits a metric of zero scalar curvature, and let $M_2$ be {\it any} compact manifold.  If $M_1 \times M_2$ belongs to Kazdan-Warner class $Z$, then so does $M_1$.

\ss

\nt{\bf PROOF :} \ \ Clearly $M_1$ cannot belong to $N$.  Suppose that it belongs to $P$.  Let $g_1$ be a metric of positive scalar curvature on $M_1$, and let $g_2$ be any metric on $M_2$.  By means of a suitable constant re-scaling of $g_1$, we can [since $M_2$ is compact] force the Riemannian product $M_1 \times M_2$ to accept a metric of positive scalar curvature, contradicting the assumption.  Hence $M_1$ must in fact be in $Z$.

\ss

\nt{\bf COROLLARY :} \  \ If $M_1$ and $M_2$ are compact scalar-flat Riemannian manifolds, and $M_1 \times M_2$ is in $Z$, so are both $M_1$ and $M_2$.

\ss

In view of this corollary, it is extremely reasonable to suppose that $Z$ is ``closed under multiplication''.

\ms

\nt{\bf CONJECTURE (3.2) [``MULTIPLICATION'']} : \  If $M_1$ and $M_2$ are compact manifolds in Kazdan-Warner class $Z$, then $M_1 \times M_2$ is again in class $Z$.

\ss

This conjecture has been verified for all known pairs of compact manifolds in $Z$.  If both manifolds are simply connected and of dimension at least 5, the statement is certainly valid [by the theorem of Stolz [23]].  All known examples of  manifolds in $Z$ admit metrics of non-generic holonomy [7], and this is a property preserved by taking products.  In short, there is strong evidence in favour of this conjecture : we shall assume its validity henceforth.  We can now say a great deal about Kazdan-Warner class $Z$, particularly in dimensions below 8.  We shall confine our attention to spin manifolds, since these are of greatest physical interest.  [Note that non-spin manifolds can be in $Z$; the underlying manifold of an Enriques surface [24] provides an example.]

\ms

\nt{\bf THEOREM (3.3)} : \ Let $N^n$ be a compact spin manifold of dimension $n$, $2 < n < 8$.  If the multiplication conjecture (3.2) is valid, then $N^n$ belongs to Kazdan-Warner class $Z$ if and only if it is diffeomorphic to one of the following:
\ss
\i [a]/  The underlying manifold of a flat space,

\i [b]/ $K3$ [if $n = 4$],

\i [c]/  $[T^r \times K3]/F$, where $T^r$ is the $r$-torus, $n=4+r$, and $F$ is a certain finite group which acts freely on $T^r \times K3$.

\ms

\nt{\bf PROOF [Sketch]} : \ Let $N^n$ satisfy the given conditions.  A theorem of Bourguignon [see [7], page 128] implies that any scalar-flat metric on $N^n$ must in fact be Ricci-flat.  [This uses the assumption that $N^n$ is in $Z$; it is not true in general, of course.]  By the Cheeger-Gromoll theorem [see [7], page 168], the universal Riemannian cover of $N^n$ has the form $\R^r \times {\widetilde Q}^{n-r}$, where $\R^r$ has its usual flat metric and ${\widetilde Q}^{n-r}$ is compact, simply connected, and Ricci-flat.  The isometry group of $\R^r$ is of course the usual Euclidean group, a Lie group with a finite number of connected components.  The isometry group of ${\widetilde Q}^{n-r}$  is a finite group.  Regarding the fundamental group $\pi_1(N^n)$ as a group of deck transformations, we see that $\pi_1(N^n)$ can be embedded discretely in a Lie group with a finite number of connected components.  By [14], section 4, the {\it stable Gromov-Lawson-Rosenberg conjecture} is valid for $N^n$.  Let us explain this conjecture.

\ss

Lichnerowicz [see [9], page 160] showed that the ${\widehat A}$ genus of a compact Riemannian spin manifold which admits a metric of positive scalar curvature must vanish.  This result was generalised by Hitchin and subsequently by Rosenberg [see [14], section 4].  Rosenberg constructs a generalised ``Dirac operator'' on any compact spin manifold $M$, with an index $\alpha(M,f)$, where $f : M \to B\pi_1(M)$ is the classifying map for the universal covering, and $\alpha(M,f)$ takes its values in a certain $KO$ group associated with a $C^\ast$ algebra defined by $\pi_1(M)$.  The details are not essential here : the crucial points are (1) that $\alpha(M,f)$ is a Dirac index which counts harmonic spinor fields in a chirality - sensitive way [that is, harmonic spinors of opposite chiralities contribute oppositely to the count], and (2) that $\alpha(M,f)$ vanishes if $M$ admits any Riemannian metric of positive scalar curvature.  If $M$ is simply connected and of dimension at least five, then the vanishing of $\alpha(M,f)$ [known in this case as $\alpha(M)$, the Lichnerowicz-Hitchin invariant] is actually {\it sufficient} as well as necessary for the existence of a metric of positive scalar curvature [23].  [Readers who are more familiar with $\alpha(M)$ should note that, unlike $\alpha(M)$, $\alpha(M,f)$ does not automatically vanish in dimension six.]  Unfortunately, this is not always so in the non-simply-connected case, but the following statement, the stable Gromov-Lawson-Rosenberg conjecture, does hold in our case.  Let $J^8$ be a compact, simply connected, eight-dimensional spin manifold with ${\widehat A}(J^8)=1$.  If $M$ is a compact spin manifold, then $M \times J^8 \times J^8 \times ...$, for a sufficient number of $J^8$ factors, admits a metric of positive scalar curvature, if and only if $\alpha(M,f)$ is zero.  We saw earlier that in our case, the fact that $N^n$ is in Kazdan-Warner class $Z$ implies that $\pi_1(N^n)$ embeds discretely in a Lie group with a finite number of connected components, and this in turn means that the stable GLR conjecture is valid for $N^n$.

\ss

Now suppose that $\alpha(N^n,f) =0$, and consider any Joyce manifold [25], [26] of holonomy Spin(7).  Such manifolds are always compact, simply connected, eight-dimensional spin manifolds with ${\widehat A}=1$.  Hence we can take $J^8$ to be such a manifold.  A Spin(7) metric on $J^8$ is scalar-flat [in fact, Ricci-flat], so $J^8$ is not in class $N$; the fact that ${\widehat A}(J^8)=1$, combined with the Lichnerowicz theorem, means that $J^8$ is not in class $P$; hence it is in $Z$.  By Conjecture (3.2), $N^n \times J^8 \times J^8 \times ...$ is likewise in $Z$.  But we know that this manifold admits  a metric of positive scalar curvature.  The contradiction shows that $\alpha(N^n,f) \not= 0$.

\ss

An alternative formulation of the Cheeger-Gromoll theorem to the one used above [see [7], pages 168-169] allows us to express $N^n$ as
$$N^n = [T^r \times {\widehat Q}^{n-r}]/F,$$

\nt where $T^r$ is a flat torus, ${\widehat Q}^{n-r}$ is a finite quotient of ${\widetilde Q}^{n-r}$, and $F$ is a finite group acting freely and isometrically on $T^r \times {\widehat Q}^{n-r}$.  Now $\alpha(N^n,f) \not= 0$ means that $N^n$ admits a non-trivial harmonic spinor;  by the formula relating the square of the Dirac operator to the Laplacian, such a spinor must in fact be {\it parallel} [that is, have zero covariant derivative] since the scalar curvature is zero.  A theorem of Futaki [13] states that the existence of a parallel spinor on a Riemannian product implies the existence of parallel spinors on each factor; thus ${\widehat Q}^{n-r}$  has a parallel spinor, and so does each factor in the de Rham decomposition [ [7], page 288] of ${\widetilde Q}^{n-r}$, its universal cover.  [The reader should note at this point that, if the group $F$ is not trivial, the fact that $N^n$ is in $Z$ does not directly imply that ${\widehat Q}^{n-r}$ is in $Z$ : it is conceivable that all metrics of positive scalar curvature on $T^r \times {\widehat Q}^{n-r}$ might fail to project to $N^n$.  Similarly, even if ${\widehat Q}^{n-r}$ is in $Z$, this does not in itself mean that ${\widetilde Q}^{n-r}$ is in $Z$.]

\ss

As is well known, the existence of parallel spinors imposes strong conditions on the holonomy group of a manifold.  Conversely, the representation theory of the holonomy group controls the number and [in relevant cases] the chiralities of the parallel spinors on a spin manifold [27], [28].  An irreducible simply connected compact $m$-dimensional Riemannian spin manifold admitting at least one parallel spinor must have a  linear holonomy group isomorphic to $SU(m/2)$, $Sp(m/4)$, $G_2$ [if $m=7$], or Spin(7) [if $m=8$].  For $N^n$ with $2 < n < 8$, the possible structures for $N^n$ are therefore of the form $[T^r \times {\widehat Q}^{n-r}]/F$, with four candidates for ${\widehat Q}^{n-r}$, as follows:

\ss

\i [a]/ ${\widehat Q}^{n-r}$ is trivial, so that $N^n  = T^n/F$.  That is, $N^n$ is a compact, flat manifold.  Here $F$ is in fact the holonomy group [29]; the possibilities for $F$ are known explicitly in low dimensions, and in  principle they are known in all dimensions [30].

\ss

\i [b]/ $n-r=4$, with ${\widehat Q}^4$ having a holonomy group locally isomorphic to $SU(2)$.  Here we say ``locally'' because, in general, ${\widehat Q}^{n-4}$ need not be simply connected, so its holonomy group is usually a disconnected Lie group.  In this particular case, however, one can show that, among the compact four-dimensional manifolds with holonomy locally isomorphic to $SU(2)$ [that is, to be more precise, having $SU(2)$ as the connected component containing the identity element], only $K3$, endowed with a Ricci-flat Kaehler metric, is actually a {\it spin} manifold [24].  The holonomy group is then {\it precisely} $SU(2)$.  Thus $N^n$ is $K3$ itself if $n=4$, or of the form $[T^r \times K3]/F$, $r=1,2,3$, if $n=5,6,7$.  Here $F$ must act non-trivially on $T^r$; the possibilities for $F$ can be analysed using the methods of [31], [32].

\ss

\i [c]/ $n-r=6$, with ${\widehat Q}^6$ having local holonomy $SU(3)$.  If ${\widehat Q}^6$ is simply connected, then it is certainly {\it not}  in Kazdan-Warner class $Z$, because, in fact, {\it every} compact simply connected six-dimensional manifold admits a metric of positive scalar curvature [ [9], page 301].  Thus we  have $\alpha({\widehat Q}^6) = 0$ in this case.  To understand this fact from the point of view of parallel spinors, note that Wang [27] shows, {\it using only the representation theory of} $SU(m/2)$, that, when $m/2$ is odd, a manifold of holonomy $SU(m/2)$ has two independent parallel spinors of {\it opposite} chiralities.  [Actually, to be precise, we should distinguish the {\it spin} holonomy group from the {\it linear} holonomy group [24].  For these manifolds, the two groups coincide, provided that the spin structure is chosen appropriately; we shall always assume that this choice has been made here.]  Since the index counts spinors of opposite chiralities in a graded way, we again have $\alpha({\widehat Q}^6) = 0$.  This derivation has the advantage of showing that the index vanishes as long as the holonomy group of ${\widehat Q}^6$ is precisely $SU(3)$.  But this is in fact so, whether or not ${\widehat Q}^6$ is simply connected, provided it is spin [in fact, provided that it is orientable].  That is, the holonomy group of a Calabi-Yau manifold is {\it precisely} $SU(3)$ in six dimensions, whether or not the manifold is simply connected.   [This is a consequence of the holomorphic Lefschetz fixed-point theorem [see [33]]: a holomorphic map has to leave the holomorphic 3-form invariant if it is to act freely.  The same argument shows that, by contrast, the quotient of an 8-dimensional Calabi-Yau manifold by a non-trivial group of isometries {\it never} has a connected holonomy group.]  Thus in fact $\alpha({\widehat Q}^6,f)=0$ whether or not ${\widehat Q}^6$ is simply connected.  This shows that ${\widehat Q}^6$ itself cannot be in class $Z$.  The same is therefore true of $T^1 \times 
{\widehat Q}^6$, where $T^1$ is of course the circle.  The only remaining possibility in this case is that $[T^1 \times {\widehat Q}^6]/F$ could be in $Z$ for some non-trivial $F$.  In fact the structure of the isometry group of ${\widehat Q}^6$ is such [32] that the only way to construct an example of this kind is as follows.  Suppose that ${\widehat Q}^6$ admits an antiholomorphic involution $\sigma$ with no fixed point [34]; we can choose an $SU(3)$ metric such that $\sigma$ acts isometrically.  The quotient has holonomy $SU(3) \triangleleft \Z_2$ [and is non-orientable, so the change in the holonomy group does not contradict our earlier assertion that taking quotients does not affect the holonomy group in the {\it spin} case; here $\triangleleft$ is the semi-direct product.].  Now regard $T^1$ as the set of unimodular complex numbers, and define an orientation-preserving involution on $T^1 \times {\widehat Q}^6$ by
$$\Sigma : (t,q) \to ({\overline t}, \sigma(q)).$$

\nt The orientable quotient $[T^1 \times {\widehat Q}^6]/ \Z_2$, with $\Z_2$ generated by $\Sigma$, has a holonomy group $SU(3) \triangleleft \Z_2$, with $\Z_2$ generated by the $SO(7)$ matrix diag $(1, 1, 1, ^-1, ^-1, ^-1, ^-1)$, and so $SU(3) \triangleleft \Z_2$ is indeed a subgroup of $SO(7)$.  In fact, it is a subgroup of $G_2$, as one can see as follows.  Let $T^1$ be parametrised by $\theta$, let $\phi$ be the Kaehler form on ${\widehat Q}^6$, and let $\omega$ be the holomorphic 3-form on ${\widehat Q}^6$.  Then [see, for example, [25], page 272] the 3-form
$$d\theta \wedge \phi + Re \; \omega,$$

\nt where $Re \ \omega$ denotes the real part of $\omega$, is a parallel 3-form defining a $G_2$ structure on $T^1 \times {\widehat Q}^6$.  Of course, $d\theta$ is not invariant under complex conjugation, and neither $\omega$ nor $\phi$ is invariant with respect to $\sigma$; but the 3-form is invariant with respect to $\Sigma$, so it descends to $[T^1 \times {\widehat Q}^6]/ \Z_2$, which means that the latter is a $G_2$ manifold.  [That is, $SU(3) \triangleleft \Z_2$ is contained in $G_2$.]  Translating this discussion into the language of parallel spinors [27], none of the parallel spinors on $T^1 \times {\widehat Q}^6$ {\it individually} survives the projection, but one particular combination does, so that, like any manifold of $G_2$ holonomy, $[T^1 \times {\widehat Q}^6]/  \Z_2$ has {\it one} parallel spinor, corresponding to the $G_2$ 3-form.  This case can therefore be treated in the next section.

\ms

\i [d]/ $n-r=7$, so $N^7 = {\widehat Q}^7$, a manifold of holonomy $G_2$.  Here again the holonomy group is independent of the fundamental group [33]; there is one parallel spinor [28], whether or not the manifold is simply connected.  The obstruction to the existence of a positive scalar curvature metric is therefore the same in all cases, namely zero. [In fact, since the fundamental group is finite in this case, the vanishing of the index here follows from general considerations [21].]  Hence, since we saw that the multiplication conjecture implies a non-zero index for all members of Kazdan-Warner class $Z$, we conclude that none of the manifolds in [d] or [c] is in fact in $Z$.  This leaves us with [a] and [b], as claimed.

\ms

Conversely, it is clear that no compact flat manifold can be in class $N$.  Nor can it be in $P$, by [9], page 306.  Hence such manifolds are in $Z$.  Again, $K3$ admits a Ricci-flat metric, so it is not in $N$; on the other hand, ${\widehat A}(K3) = 2$, so it is not in $P$; thus it is in $Z$.  It can be shown that these facts imply that the $\alpha$-invariant of any manifold of the form $[T^r \times K3]/F$ is non-zero, so again these manifolds are in $Z$.  This completes the proof of Theorem (3.3).

\ms

The point of this theorem is that it allows us to identify the members of class $Z$ [in dimensions below 8] very explicitly.  The question of the existence of a stable CFT is thereby reduced to a study of the problem for these specific manifolds.  We shall discuss this in the next section.

\ms

Compact six-dimensional Riemannian manifolds of holonomy $SU(3)$ are, according to Theorem (3.3), all in Kazdan-Warner class $P$, so a stable Maldacena CFT exists on all of them.  By contrast, the two parallel spinors on a compact eight-dimensional manifold of holonomy $SU(4)$ are of the {\it same} chirality [27] and so ${\widehat A} = 2$ for these manifolds: hence, all of them are in class $Z$.  Since there is no known classification of Calabi-Yau manifolds in any dimension above four, we see that there is no hope of extending the explicit characterisations of Theorem (3.3) beyond $n=7$.  A broader classification, suited to producing examples, can however be given : we shall conclude this section with a sketch of it.

\ms 

Assuming the validity of the multiplication conjecture, we find that the only examples of compact nine and ten dimensional manifolds in $Z$ are [global or local] products of tori with lower-dimensional manifolds.   The real novelties are in dimension eight, so we concentrate on that case.  The most useful classification is in terms of holonomy groups.  Note that we discuss here only {\it spin} manifolds; there are many other eight-dimensional compact manifolds in $Z$ with other holonomy groups [[for example, all compact eight-dimensional manifolds of holonomy $[\Z_8 \times SU(4)]/ \Z_4$ are in $Z$] but they are {\it not} spin [24].

\ms

A compact eight-dimensional spin manifold in Kazdan-Warner class $Z$ belongs to one of the following classes, if the multiplication conjecture is valid :

\ss

\i [i]/ Flat manifolds $T^8/F$, where $F$ is finite and is isomorphic to the holonomy group.

\ss

\i [ii]/ Local products of lower-dimensional Ricci-flat manifolds with tori, such as $[T^4 \times K3]/F$.  Note that, if $T^2$ is  the orientable double cover of a Klein bottle, and ${\widehat Q}^6$ is a six-dimensional Calabi-Yau manifold which admits an antiholomorphic involution, then $[T^2 \times {\widehat Q}^6]/ \Z_2$ [with $\Z_2$ acting non-trivially on both factors] is in $Z$, even though $T^2 \times {\widehat Q}^6$ itself is in $P$.  This manifold has holonomy $SU(3) \triangleleft \Z_2$, a subgroup of $SU(4) \triangleleft \Z_2$ [see below].

\ss

\i [iii]/ Manifolds of holonomy $SU(2) \times SU(2)$, $[\Z_4 \times SU(2) \times SU(2)]/ \Z_2$, $[Q_8 \times SU(2) \times SU(2)]/ \Z_2$, and $[SU(2) \times SU(2)] {\widetilde \triangleleft} \ Z_8$.  Here $Q_8$ is the quaternionic group of order 8, and the notation ${\widetilde \triangleleft}$ is intended to convey the idea that $\Z_8$ acts on $SU(2) \times SU(2)$ but has a non-trivial intersection with it, so the product is not semi-direct in the strict sense.  Of course, $SU(2) \times SU(2)$ is the holonomy group of $K3 \times K3$, where $K3$ has been assigned a Ricci-flat Kaehler metric.  The quotient of $K3 \times K3$ by the $\Z_2$ generated by a fixed-point-free holomorphic involution $T_+$ yields a manifold of holonomy $[\Z_4 \times SU(2) \times SU(2)]/  \Z_2$, where $SU(2) \times SU(2)$ should be regarded as a subgroup of $SU(4)$, in the obvious way, and where $\Z_4$ is generated by the $SU(4)$ matrix $iI_4$.  Such a $T_+$ may have a ``square root'' of order four, acting holomorphically and without fixed point on $K3 \times K3$, and the quotient then has holonomy $[SU(2) \times SU(2)] {\widetilde \triangleleft} \ Z_8$, with $SU(2) \times SU(2)$ embedded in $SU(4)$ as above, and $\Z_8$ generated by the $U(4)$ matrix [with $\sqrt{i}$ denoting either square root of $i$]

$$z = \left[ \matrix{ 0                  &\sqrt{i} \ I_2 \cr
                             \sqrt{i} \ I_2  &0 \cr} \right].$$

\i / Note that conjugation by $z$ induces the exchange automorphism on $SU(2) \times SU(2)$, that $z^2 = i \ I_4$ [which is not in $SU(2) \times SU(2)$], and that $z^4 = -I_4$, which is in $SU(2) \times SU(2)$, so that $[SU(2) \times SU(2)] {\widetilde \triangleleft} \ Z_8$ has four [not eight] connected components.  Finally, suppose that $K3 \times K3$ admits an antiholomorphic involution $T_-$ which has no fixed point and which commutes with $T_+$.  The quotient of $K3 \times K3$ by the $\Z_2 \times \Z_2$ group generated by $T_+$ and $T_-$ then has holonomy $[Q_8 \times SU(2) \times SU(2)] / \Z_2$.  Here one should think of $SU(2)$ as the symplectic group $Sp(1)$, so that $Sp(1) \times Sp(1)$ is a subgroup of $Sp(2)$; $Q_8$ is a subgroup of $Sp(1)$ in an obvious way, so that $[Q_8 \times SU(2) \times SU(2)] / \Z_2$ is contained in the canonical $[Sp(1) \times Sp(2)/ \Z_2$ subgroup of $SO(8)$; in other words, $[K3 \times K3] / [\Z_2 \times Z_2]$ is a quaternion-Kaehler manifold [7].  Examples of all these kinds can be constructed by using the particular representation of $K3$ given in [35].  We shall present the details elsewhere.
Further examples of manifolds of holonomy $[Q_8 \times SU(2) \times SU(2)]/ \Z_2$ [with this group given as $\Z^2_2 \triangleright SU(2)^2]$ may be found in [25], page 390.  Note that $[Q_8 \times SU(2) \times SU(2)]/  \Z_2$ is in fact a subgroup of $SU(4) \triangleleft \Z_2$.

\ss

\i [iv]/ Manifolds of holonomy $Spin(5)$ [ = $Sp(2)$] and $\Z_3 \times Spin(5)$.  These are the 8-dimensional hyperKaehler manifolds and their quotients by the group generated by a fixed-point-free holomorphic map of order three.  For a discussion of examples with holonomy $Spin(5)$, see [25], page 166; for the $\Z_3 \times Spin(5)$ case, see [34]; no example has yet been found in this second, equally interesting case.

\ms

\i [v]/ Manifolds of holonomy $Spin(6)$ [=$SU(4)$] and $Spin(6) \triangleleft \Z_2$.  These are the 8-dimensional Calabi-Yau manifolds and their quotients by the group generated by a fixed-point-free antiholomorphic involution. Compact 8-dimensional Calabi-Yau manifolds are, unlike their six-dimensional counterparts, necessarily simply connected; otherwise, examples can be constructed in the same way.  Their quotients by a $\Z_2$ generated by an antiholomorphic involution are non-Kaehler spin manifolds [24] of holonomy $Spin(6) \triangleleft \Z_2$.  Examples may be found in [34]; further examples have been constructed in [25], page 400.

\ms

\i [vi]/ Manifolds of holonomy $Spin(7)$.  These have one parallel spinor, are necessarily simply connected, and admit no quotients.  Many examples are given in [25].  Note that all of the manifolds discussed in [iii], [iv], [v] above are $Spin(7)$ manifolds, in the sense that their holonomy groups are contained in $Spin(7)$.  In a sense, therefore, one can say that {\it a ``generic'' eight-dimensional compact manifold in Kazdan-Warner class} $Z$ {\it is a manifold admitting a metric of holonomy} $Spin(7)$.  From the point of view of the stability of Maldacena-type conformal field theories, therefore, $Spin(7)$ manifolds are of basic importance; they play a role in eight dimensions analogous to that played by $K3$ in four dimensions.  The great difference is that $K3$ is the unique spin non-flat four-dimensional member of $Z$, while the number of distinct $Spin(7)$ manifolds is thought [ [25], page 417] to be at least in the tens of thousands.

\ms

This completes our survey of Kazdan-Warner class $Z$.  We now discuss the physical significance of our findings.

\bs

\nt{\bf 4. \ DISCUSSION.}
\ss
The $AdS/CFT$ correspondence ``replaces'' [36] space-time geometry and physics with a simpler {\it conformal} theory on a {\it conformal} manifold.  Because of its role in ensuring conformal coupling of scalar fields, the scalar curvature of ``infinity'' is often of great importance in the $AdS/CFT$ correspondence and its generalisations.  For example, the familiar ${\cal N}  = 4$ superconformal gauge theory on $S^4$ has a moduli space described by the expectation values of six scalar fields $X^a$ governed by an action of the form
$$\int d^4 x \sqrt{g} [Tr(dX)^2 + {1\over 6} R \; Tr X^2],$$

\nt where $R$ is the scalar curvature.  If $S^4$ is given a metric of positive scalar curvature, then [4] it is the term involving $R$ that ensures the convergence of the path integral for the partition function.  Again, the fact that $R$ appears in the conformal Laplacian means that positive $R$ ensures that the $BPS$ brane action in the bulk is bounded below [3], [5].  Positive $R$ again appears when one considers gauge symmetry breaking in the $AdS/CFT$ context [37], since it is evident from the action above that a constant set of $X^a$ is not possible when $R$ is positive: so positive $R$ obstructs symmetry-breaking.

\ss

In all these cases, the role of {\it negative} scalar curvature is rather  clear $-$ as one would expect, it has the opposite effect to positive scalar curvature.  By contrast, the role of  {\it zero} scalar curvature is ambiguous; yet it could be of great importance.  For example, a zero scalar curvature metric on $S^4$ [we remind the reader that {\it such a metric certainly exists}, though its precise form is not known] would permit non-trivial expectation values for the scalars $X^a$, without necessarily causing instability.  Clearly, then, the study of manifolds of zero scalar curvature is potentially of great interest in string theory.

\ss

The most basic question in this regard is the following: does a given manifold $N^n$ admit {\it any} metric such that the $CFT$ is well-behaved with respect to that metric?  The Kazdan-Warner classification immediately yields a partial answer : ``yes'' for class $P$, ``no'' for class $N$.  There remains the mysterious class $Z$.  Recall that these are manifolds such that a smooth function can be a scalar curvature of some  metric, provided that the function is either negative somewhere or identically zero.  [By contrast, {\it any} smooth function can be a scalar curvature on a manifold in class $P$, such as the sphere $S^4$; we must resist the intuitive association of spheres with positive scalar curvature.]  These manifolds admit {\it no} metric of positive scalar curvature : understanding them is the key to a complete understanding of stability questions in the $AdS/CFT$ context, as well as of closely related issues [such as symmetry breaking] connected with conformally coupled scalar fields.

\ms

Fortunately, Theorem (3.3) implies that manifolds in $Z$ are, in dimensions below 8, both
very rare and very familiar. In fact, conformal field theories of the relevant kind have
already been studied on these manifolds or finite covers of them.

\ms

The tori $T^r$ are, of course, boundaries of certain manifolds-with-boundaries, and $AdS/CFT$
was studied in this context in [38]. The conformal field theory on $T^r$ is treated by 
imposing the appropriate periodic and antiperiodic boundary conditions on fields defined on
$\R^r$, the universal cover. The corresponding supergravity theory in the bulk has some 
unusual properties $-$ the bulk metric, given by equation (50) of [38], is apparently
singular $-$ but the effects of compactification on the conformal field theory are innocuous:
there are contributions to the Casimir energy density, but no suggestion that the theory has
become unstable. Field theories on other flat manifolds can be treated as follows. Any such
manifold is obtained, as we know, by taking the quotient of $T^r$ by a certain {\it finite}
group of isometries, $F$. Because $F$ is finite, we can always construct $F$-invariant objects
on $T^r$ by an averaging process, and these averaged objects project to $T^r/F$. Since $F$
is isometric, the flat metric also projects, of course. Thus we see that the theory on
the quotient is no more susceptible to instability than its counterpart on $T^r$. Notice
the importance of the finiteness of $F$ in Theorem (3.3). Notice too that this situation
is not like the one we discussed in Section 2 in connection with the manifold 
$[\R P^7 \times S^2] \# [{\hbox{~exotic~}} S^9$]. A $CFT$ on that manifold is unstable 
because the metrics of positive and zero scalar curvature on its double cover {\it do not}
project to the quotient.

\ms

The $AdS/CFT$ correspondence for field theories on $K3$ does not appear to have been
investigated thus far. [The fact that $K3$ is not a boundary should be no impediment;
see [4], [16].] However, the relevant kind of $CFT$ was explored in great depth in [39].
A partition function for the $SU(2)$ theory was explicitly written down [equation (4.17)
of [39]] and a similar formula was conjectured for the $SU(N)$ theory on $K3$. There is
no evidence of instability in any of the results of [39]. Hence we can be confident that
there are stable field theories on $T^r \times K3$; then, as above, CFT configurations
on manifolds of the form $[T^r \times K3]/F$ can be thought of as $T^r \times K3$ 
configurations respecting the finite symmetry group $F$, and so this settles all cases
in Theorem (3.3). 

\ms

Assuming the validity of the multiplication and $AdS/CFT$ conjectures, then, we can
draw the following conclusion. For values of $n$ less than eight, string backgrounds
which can be expressed as a connected component of the complement [in a compact 
10-dimensional manifold] of a hypersurface of the form $N^n \times Y^{9-n}$ are
intrinsically [that is, topologically] subject to Seiberg-Witten brane instability
{\it if and only if} $N^n$ belongs to Kazdan-Warner class $N$.

\ms

For $N^n$ of dimension eight [and therefore in higher dimensions] the situation is dramatically different : there is certainly no theorem remotely as precise as Theorem (3.3).  Compact eight-dimensional manifolds of holonomy $Spin(5)$ and $\Z_3 \times {\rm Spin}(5)$ seem to be rather rare, but there are multitudes of manifolds of holonomy $Spin(6)$, $Spin(6) \triangleleft \Z_2$, and $Spin(7)$.  All of these are in class $Z$.  The $Spin(7)$ case promises to be particularly interesting.  The very much simpler Kaehler manifold discussed earlier, $[K3 \times K3] / \Z_4$,  has much in common with manifolds of holonomy $Spin(7)$; perhaps a study of conformal field theory on it will be useful in this respect.

\ms

Because we have concentrated on low-dimensional examples, the question of the stability of conformal field theories on manifolds with exotic differentiable structures has not been a major issue in this work.  For reasons explained in the Introduction, however, we are convinced that this will be an important complication for $AdS/CFT$ when the $CFT$ is defined on a manifold of eight or more dimensions.  Section (3.2) of [40] is undoubtedly relevant here.

\ss

The full $AdS/CFT$ correspondence [see [41], Chapter 3] potentially applies
to generalised $AdS$ bulks containing a wide variety of objects: strings,
D-branes, black holes, and so on. Work is currently in progress on applying
the topological stability criterion to physically interesting backgrounds
of this kind; let us conclude with some brief comments on some of these
applications. The brane case has been discussed, in the Randall-Sundrum
context, in [17]; there the stability criterion appears as a constraint
on the sign of the cosmological constant. We shall return to this 
application in a forthcoming work. The possibility of extending $AdS/CFT$
to Ricci-flat infinities, as discussed here, was considered in [42].
Our stability criterion guarantees that the very delicate test of the
correspondence in [42] will not be disturbed by instabilities. Finally,
the criterion also has a very interesting application in the study of
conformal field theories at finite temperatures. The $AdS$ duals in this
case are black holes. The horizons of asymptotically flat black holes have very restricted topologies, but there are many more possibilities in the
asymptotically $AdS$ case [43]. In particular, one has the five-dimensional
asymptotically $AdS$ black hole metric

$$g^W = -f(r) \; dt \otimes dt + f^{-1}(r) \; dr \otimes dr + r^2h_{ij} \; dx^i \otimes dx^j$$

with

$$f = -1 - {16\pi G \mu \over 3r^2 \; {\rm{Vol}}(M^3)} + {r^2 \over L^2}$$

where ${\mu}$ is the mass parameter and $h$ is the metric on the 
three-dimensional horizon, $M^3$. This is a solution of the Einstein 
equations provided that $M^3$ is a compact three-manifold of {\it constant
negative curvature}. The conformal boundary of the Euclidean version of
this space is just $S^1 \times M^3$; this is the negative-curvature 
analogue of the black hole solution considered by Witten in [4] and [44].
There the boundary was $S^1 \times S^3$, and the Hawking-Page [45] phase
transition in the bulk was interpreted as a confinement/deconfinement
transition in the CFT on $S^1 \times S^3$. This argument has been extended
recently [46] to the case of $AdS$ black holes with compact {\it flat}
horizons; this is done by using the Horowitz-Myers $AdS$ soliton [47] as thermal background. It is pointed out in [46] that there is apparently
no analogue of the Horowitz-Myers soliton for the hyperbolic black hole 
given above, and so it is unclear whether the thermodynamics of that
case can be understood in a similar way. Now $S^1 \times M^3$ evidently
has negative scalar curvature, so the full string theory on this background
is unstable to the emission of large branes. Furthermore, it is possible
to show that $S^1 \times M^3$ is in Kazdan-Warner class $N$, so that,
even if one were to attempt to take any back-reaction into account,
the instability cannot be avoided: we have another example of topologically
induced instability. This is undoubtedly relevant to the difficulty,
observed in [46], of studying the thermodynamics of the hyperbolic black
hole in the $AdS/CFT$ context. These and other applications of the 
topological instability criterion will be discussed elsewhere.

\bs
\bs

\nt{\bf REFERENCES}
\ss
\i [1]/ R. P. Geroch and G. T. Horowitz, in General Relativity : an Einstein Centenary Survey, eds S. W. Hawking and W. Israel, Cambridge University Press, 1979.

\i [2]/ M. Lachieze-Rey and J. P. Luminet, Cosmic Topology, Phys. Rept. {\bf 254} (1995) 135 [gr-qc/9605010].

\i [3]/ N. Seiberg and E. Witten, The D1/D5 System and Singular CFT, JHEP {\bf 9904} (1999) 017 [hep-th/9903224].

\i [4]/ E. Witten, Anti de Sitter Space and Holography, Adv. Theor. Math. Phys. {\bf 2} (1998) 253 [hep-th/9802150].

\i [5]/ E. Witten and S. T. Yau, Connectedness of the Boundary in the AdS/CFT Correspondence [hep-th/9910245].

\i [6]/ E. Witten, http://online.itp.ucsb.edu/online/susy$_{-}$c99/witten/

\i [7]/ A. L. Besse, Einstein Manifolds, Springer Verlag, 1987.

\i [8]/ J. G. Ratcliffe, Foundations of Hyperbolic Manifolds, Springer Verlag, 1994.

\i [9]/ H. B. Lawson and M. L. Michelsohn, Spin Geometry, Princeton University Press, 1989.

\i [10]/ A. Kehagias and J. G. Russo, Hyperbolic Spaces in String and M Theory, JHEP {\bf 0007} (2000) 027 [hep-th/0003281].

\i [11]/  G. D. Starkman, D. Stojkovic, and M. Trodden, Large Extra Dimensions and Cosmological Problems [hep-th/0012226].

\i [12]/ J. L. Kazdan and F. W. Warner, A Direct Approach to the Determination of Gaussian and Scalar Curvature Functions, Invent. Math. {\bf 28} (1975) 227.

\i [13]/ A. Futaki, Scalar-Flat Closed Manifolds Not Admitting Positive Scalar Curvature Metrics, Invent. Math. {\bf 112} (1993) 23.

\i [14]/ J. Rosenberg and S. Stolz, Metrics of Positive Scalar Curvature and Connections with Surgery, http://www.math.umd.edu/$\sim$jmr/jmr$_{-}$pub.html

\i [15]/ R. Schoen, Conformal Deformation of a Riemannian Metric to Constant Scalar Curvature, J. Differential Geom. {\bf 20} (1984) 479.

\i [16]/ B. McInnes, AdS/CFT for Non-Boundary Manifolds, JHEP {\bf 0005} (2000) 025 [hep-th/0003291].

\i [17]/ B. McInnes, The Topology of the AdS/CFT/Randall-Sundrum Complementarity [hep-th/0009087].

\i [18]/ R. Mazzeo, The Hodge Cohomology of a Conformally Compact Metric, J. Differential Geom. {\bf 28} (1988) 309.

\i [19]/ S. S. Gubser, Curvature Singularities : the Good, the Bad, and the Naked [hep-th/0002160].

\i [20]/ J. W. Milnor and J. D. Stasheff, Characteristic Classes, Princeton University Press, 1974.

\i [21]/ B. Botvinnik and B. McInnes, On Rigidly Scalar-Flat Manifolds [math.DG/9911023].

\i [22]/ A. Dessai, On the Topology of Scalar-Flat Manifolds [math.DG/0006149].

\i [23]/ S. Stolz, Simply Connected Manifolds of Positive Scalar Curvature, Ann. Math. {\bf 136} (1992) 511.

\i [24]/ B. McInnes, Spin Holonomy of Einstein Manifolds, Comm. Math. Phys. {\bf 203} (1999) 349.

\i [25]/ D. Joyce, Compact Manifolds with Special Holonomy, Oxford University Press, 2000.

\i [26]/ D. Joyce, A New Construction of Compact 8-Manifolds with Holonomy Spin(7), J. Differential Geom. {\bf 53} (1999) 89 [math.DG/9910002].

\i [27]/ M. Y. Wang, Parallel Spinors and Parallel Forms, Ann. Global. Anal. Geom. {\bf 7} (1989) 59.

\i [28]/ M. Y. Wang, On Non-Simply-Connected Manifolds with Non-Trivial Parallel Spinors, Ann. Global. Anal. Geom. {\bf 13} (1995) 31.

\i [29]/ J. A. Wolf, Spaces of Constant Curvature, Publish or Perish, 1974.

\i [30]/ L. S. Charlap, Bieberbach Groups and Flat Manifolds, Springer Verlag, 1986.

\i [31]/ S. Mukai, Finite Groups of Automorphisms of K3 Surfaces and the Mathieu Group, Invent. Math. {\bf 94} (1988) 183.

\i [32]/ B. McInnes, The Quotient Construction for a Class of Compact Einstein Manifolds, Comm. Math. Phys. {\bf 154} (1993) 307.

\i [33]/ B. McInnes, Holonomy Groups of Compact Riemannian Manifolds : a Classification in Dimensions up to Ten, J. Math. Phys. {\bf 34} (1993) 4273.

\i [34]/ B. McInnes, Examples of Einstein Manifolds with All Possible Holonomy Groups in Dimensions Less Than Seven, J. Math. Phys. {\bf 34} (1993) 4287.

\i [35]/ N. Hitchin, Compact Four-Dimensional Einstein Manifolds, J. Differential Geom. {\bf 9} (1974) 435.

\i [36]/ G. T. Horowitz, Quantum Gravity at the Turn of the Millennium, [gr-qc/0011089].

\i [37]/ I. R. Klebanov and E. Witten, AdS/CFT Correspondence and Symmetry Breaking, Nucl. Phys. {\bf B556} (1999) 89 [hep-th/9905104].

\i [38]/ R. C. Myers, Stress Tensors and Casimir Energies in the AdS/CFT Correspondence Phys. Rev. {\bf D60} (1999) 046002 [hep-th/9903203].

\i [39]/ C. Vafa and E. Witten, A Strong Coupling Test of S-Duality, Nucl. Phys. {\bf B431} (1994) 3 [hep-th/9408074].

\i [40]/ D. R. Morrison and M. R. Plesser, Non-Spherical Horizons I, Adv. Theor. Math. Phys. {\bf 3} (1999) 1 [hep-th/9810201]

\i [41]/ O. Aharony, S.S. Gubser, J. Maldacena, H. Ooguri, Y. Oz, Large N Field Theories, String Theory and Gravity, Phys.Rept. 323 (2000) 183 [hep-th/9905111]

\i [42]/ P. Mansfield and D. Nolland, Order $1/N^2$ test of the Maldacena conjecture: Cancellation of the one-loop Weyl anomaly, Phys.Lett. B495 (2000) 435 [hep-th/0005224]

\i [43]/ D. Birmingham, Topological Black Holes in Anti-de Sitter Space, Class.Quant.Grav. 16 (1999) 1197 [hep-th/9808032]

\i [44]/ E.Witten, Anti-de Sitter Space, Thermal Phase Transition, And Confinement In Gauge Theories, Adv.Theor.Math.Phys. 2 (1998) 505 [hep-th/9803131]

\i [45]/ S.W. Hawking and D.N. Page, Thermodynamics of black holes in anti-de Sitter space, Comm. Math. Phys. 87 (1983) 577.

\i [46]/ S. Surya, K. Schleich, D. M. Witt, Phase Transitions for Flat AdS Black Holes
 [hep-th/0101134]

\i [47]/ G. T. Horowitz and R. C. Myers, The AdS/CFT Correspondence and a New Positive Energy Conjecture for General Relativity, Phys.Rev. D59 (1999) 026005 [hep-th/9808079]

\bye